\documentclass[review]{elsarticle}

\usepackage{graphicx}
\usepackage{subfigure}
\usepackage{epsfig}
\usepackage{epstopdf}
\usepackage{color}
\usepackage{amsmath}

\usepackage{amssymb}
\usepackage{array}
\usepackage{graphicx}

\def\beq{\begin{equation}}
\def\eeq{\end{equation}}
\def\bea{\begin{eqnarray}}
\def\eea{\end{eqnarray}}

\begin{document}

\begin{frontmatter}

\title{Generating odd-dimensional rotating black holes with equal angular momenta by using the Kerr--Schild Cartesian form of metric}

\author{Masoumeh Tavakoli$^{1,2}$}
\ead[url]{m.tavakoly@ph.iut.ac.ir}


\author{Behrouz Mirza$^1$}
\ead[url]{b.mirza@iut.ac.ir}

\address{$^1$Department of Physics, Isfahan University of Technology, Isfahan 84156-83111,  Iran. \\ $^2$Research Institute for Astronomy and Astrophysics of Maragha (RIAAM), Maragha, P. O. Box: 55134-441, Iran.}

\begin{abstract}
The Newman--Janis (NJ) method is a prescription to derive the Kerr space-time from the Schwarzschild metric.
The BTZ, Kerr and five-dimensional Myers--Perry (MP) black hole solutions have already been generated by different versions of the NJ method.
 However, it is not known how to generate the metric of higher-dimensional ($d > 6$) rotating black holes by this method.\\
 In this paper, we  propose the simplest algorithm for generation of  the five-dimensional MP metric with two arbitrary  angular momenta by using the Kerr--Schild form of the metric and quaternions.
Then,  we present another 
 new two-step version of the NJ approach  without using quaternions that generate the five-dimensional  MP metric with equal angular momenta.
Finally, the extension  of  the later procedure
 is explained for  the higher odd-dimensional rotating black holes ($d > 5$) with equal angular momenta.
\end{abstract}
\end{frontmatter}
\section{Introduction}
\label{sec:intro}
Newman and Janis proposed
a procedure for generating rotating metrics (axisymmetric) from static ones (spherically symmetric)  in 1965 \cite{1,2}.
 The original Newman--Janis (NJ) trick devised for the Kerr metric  included four steps
  as follows:
First, inverting the metric to Eddington--Finkelstein coordinates, second, finding a null tetrad basis,  then, applying complex conjugate transformations  for $r$ and $u$ components  and Finally, inverting again the metric to earlier coordinates  followed by changing  to the  Boyer--Lindquist coordinates.
 In 1990, another version of the NJ trick  was developed by Giampieri, where the  null tetrads were avoided, thus facilitate the generalization of  the procedure to more complicated cases \cite{Giampieri1990}.  Recently, a simple version of the NJ method  was proposed in \cite{Visser1,Visser2}, where the authors  used both  oblate spheroidal and Cartesian coordinates to generate the four-dimensional Kerr--Schild Cartesian form of the Kerr metric. \\
 The five-dimensional Myers--Perry (MP) black hole solution was obtained
  using  simplification of the NJ algorithm proposed by Giampieri \cite{3,5,6,4,Sharifi}.
The authors in \cite{Sharifi} also proposed a simpler
NJ algorithm to generate five-dimensional rotating black holes.
  An interesting aspect of this  approach was the use of quaternions for the first time.\\
   In this paper, we  propose the simplest two steps formulation of the NJ method to obtain the five-dimensional  MP rotating black holes with
  two independent 
 angular momenta instead of using the original 4-step NJ prescription. In our approach, we have used both quaternions and the Kerr--Schild form of the metric.
  However, the extension of  this protocol to higher dimensional rotating black holes is not possible. 
  Therefore, we introduce another prescription based on the Kerr--Schild form of the metric to obtain   higher dimensional rotating black holes with equal angular momenta. \\
 Recently, an interesting application of the NJ method for calculating self force  in four and five-dimensional rotating black holes was proposed in \cite{1os,2os}.
\\
This paper is organized as follows:
In Section \ref{sec:3}, we introduce a new approach where  the Kerr--Schild form of  the five-dimensional Schwarzschild black holes  and quaternions are used to generate five-dimensional MP black holes with two angular momenta.
In Section \ref{4}, we further argue about the distance parameter replacement in our prescription by using rotation and translation of resource in the potential.
In Section \ref{AppendixA},  we introduce another  version of the NJ algorithm and the Kerr--Schild form of the metric is generated  for higher odd-dimensional rotating black holes with equal angular momenta.
In Appendix A, the basic  mathematical properties of quaternions are reviewed.
In Appendix B, we explain how to generate the Kerr--Schild Cartesian  form of  the rotating BTZ black hole solution by the simplest version of the NJ method.
\section{Rotating five-dimensional black holes}\label{sec:3}
In this Section, we
use the Kerr--Schild form of the metric and quaternions  to
obtain five-dimensional MP black holes, for earlier studies that other versions of the NJ method were introduced see \cite{4} and \cite{Sharifi}. 
In \cite{Sharifi}, the authors  proposed a new algorithm  for  generating five-dimensional Meyers-Perry black holes using quaternions.
This method was simpler to use compared to the earlier proposals in \cite{4}.
\\ We  introduce a new two-step method to generate  five-dimensional black holes.
In the Cartesian coordinates, a five-dimensional Schwarzschild metric is as follows:
\begin{eqnarray}\label{Cartesian coordinates Schwarzschild}
ds^2=-d\tau ^2+ \sum_{i=1}^{2} \Big[d x _{i}^2+d y _{i}^2\Big]+
(1-f(r_0)) \left( \sum_{i=1}^{2} \Big[\frac{(x _{i} d x _{i}  +y _{i}  d y _{i} )}{r_0}\Big]
+d\tau \right)^2,
\end{eqnarray}
where
\begin{equation}
f(r_0)=1-\frac{m}{r_0^2}.
\end{equation}
Therefore, the five-dimensional Schwarzschild geometry  in the Kerr--Schild form can be written as follows:
\begin{equation}\label{noacc5}
\mathsf{g}_{a b}=\eta_{a b}+\big(1- f(r_0)\big) (\ell_0)_a (\ell_0)_b;
\end{equation}
where
\begin{eqnarray}
\label{300}
\eta  _{a b}=\mathit{diag} \lbrace -1,+1,+1,+1,+1 \rbrace,
\end{eqnarray}
\begin{eqnarray}
\label{300}
&&(\ell_0)_a=(1;\hat{r_0})
\\
&&~~~~~~=(1;\frac{ x _{1}}{r_0},\frac{y _{1}}{r _0},\frac{x_{2}}{r _0},\frac{y_{2}}{r_0}).
\end{eqnarray}
\noindent We  define the following   coordinate transformations:
\begin{equation}
\label{1}
x _{1}+\hat{i} y _{1} =r_0 \sin(\theta_0)e^{\hat{i} \phi_{01}},~~~~~~~x _{2}+\hat{j} y _{2} =r_0 \cos(\theta_0)e^{\hat{j} \phi_{02}},
\end{equation}
where $\hat{i}$ and $\hat{j}$ are quaternions (Appendix A).
 Therefore the null co-vector in Eq. (\ref{300}) can be written as follows:
 \begin{equation}\label{3.7}
 (\ell_0)_a=(1;\sin \theta_0 \cos \phi_{01},\sin \theta_0 \sin \phi_{01},\cos \theta_0 \cos \phi_{02},\cos \theta_0 \sin \phi_{02}).
 \end{equation}
To obtain the  MP space-time, we initially
use the following oblate spheroidal coordinates ($r, \theta, \phi_1, \phi_2$):
\begin{equation}
 \label{2}
x _1+\hat{i}y _1 =(r+\hat{i} a_1) \sin \theta e^{\hat{i} \phi_1},~~~~~~~x _2+ \hat{j} y _2 =(r+\hat{j} a_2) \cos \theta e^{\hat{j} \phi_2}.
\end{equation}
The null co-vector  $(\ell_0)_a$ in Eq. (\ref{3.7}) can  be replaced as follows:
\begin{eqnarray}
   \label{500}
(\ell_0)_a
\longrightarrow \ell _a =(1;\sin \theta \cos \phi_1,\sin \theta \sin \phi_1,\cos \theta \cos \phi_2,\cos \theta \sin \phi_2),
\nonumber\\
=(1; \mathit{Re}\Big[\frac{x_1+\hat{i}y_1}{r+\hat{i} a_1} \Big], \mathit{I}\textit{m}\Big[\frac{x _1+\hat{i}y _1}{r+\hat{i} a_1} \Big], \mathit{R}\textit{e}\Big[\frac{x _2+\hat{j} y_2}{r+\hat{j} a_2} \Big], \mathit{I}\textit{m}\Big[\frac{x _2+\hat{j} y _2}{r+\hat{j} a_2} \Big])
\nonumber\\
=(1; \frac{x _1 r+y _1 a_1}{r^2+a_1^2} , \frac{y_1 r- a_1 x_1}{r^2+a_1^2} , \frac{x_2 r+y _2 a_2}{r^2+ a_2^2} , \frac{y _2 r- a_2 x _2 }{r^2+ a_2^2}).
\end{eqnarray}
Now, we use a prescription  based on quaternions that was introduced in  \cite{Sharifi}. According to \cite{Sharifi}, $r$ and $\bar{r}$ are defined in the following equations:
\begin{eqnarray}\label{r1}
r=r^\prime - \hat{i} \ a_1 \cos  \theta - \hat{j} \ a_2 \sin  \theta,\nonumber\\
\bar{r}=r^\prime + \hat{i} \ a_1 \cos  \theta + \hat{j} \ a_2 \sin  \theta.
\end{eqnarray}
Therefore, we use the following replacement for $1-f(r_0)$:
\begin{equation}\label{3.10}
\frac{m}{r^2_0}\longrightarrow  \frac{m}{r^2}=\frac{m}{ r \bar{r}}=\frac{m}{r^2+a_1^2 \cos ^2 \theta +a_2^2 \sin ^2 \theta}.
\end{equation}
It should be noted that quaternions are very useful for this replacement ($r_0^2\rightarrow r^2$) in our calculation. Using quaternions leads to a simplest algorithm  compared to complex numbers in \cite{4}.
\\Finally, using  replacements that defined  in Eqs. (\ref{500}) and (\ref{3.10}) in
 Eq.(\ref{noacc5}), the defining metric for the MP  space-time in the Kerr--Schild Cartesian form is obtained:
\begin{equation}\label{rotating Myers-Perry1}
\mathsf{g} _{a b}=\eta_{a b}+\frac{m}{r^2+a_1^2 \cos ^2 \theta +a_2^2 \sin ^2 \theta} \ell_a \ell_b.
\end{equation}
Using  Eqs. (\ref{500}) and (\ref{rotating Myers-Perry1}), we can write  the rotating MP
metric with two angular momenta in five dimensions in the Cartesian coordinates  as follows:
\begin{eqnarray}
 \label{rotating Myers-Perry}
 &ds^2  =-d\tau^2+\sum^{2}_{i=1}{\Big[dx_i^2+dy _i^2}\Big]+
\nonumber\\
 &(1-f(r)) \left( \sum^{2}_{i=1}\Big[{\frac{r (x _i dx _i + y _i dy _i)-a_i (x _idy _i -y _i dx _i )}{r^2 +a_i^2}}\Big]
+d\tau \right)^2,
\end{eqnarray}
where
\begin{eqnarray}
&&1-f(r)=\frac{m}{r^2+a_1^2 \cos ^2 \theta +a_2^2 \sin ^2 \theta}
\nonumber\\
&&~~~~~~~~~~=\frac{m}{r^2+a_1^2 (\frac{x_2^2+y_2^2}{r^2+a_2^2})+a_2^2 (\frac{x_1^2+y_1^2}{r^2+a_1^2})}.
\end{eqnarray}
In this way, by using a simple and quick algorithm,
 we  have obtained the rotating five-dimensional MP  metric in  the Kerr--Schild coordinates with two angular momenta.
 In the following Section, 
  the procedure for obtaining Eq. (\ref{r1}) is explained. 
\section{Transformation function using the Kerr--Schild scalar potential:}\label{4}
It was argued in \cite{Visser1} that in four dimensions, we may explain  ``$r\rightarrow r-\hat{i} a \cos \theta$" by  a displacement of the potential source of the  Schwarzschild  solution   in the complex plane (displacement of $z$ to $z-\hat{ i}a$) (see also \cite{Nilton, Schiffer}).\\
In five dimensions, for obtaining ``$r\rightarrow r-\hat{i} a \cos \theta - \hat{j} b \sin \theta$" we have  to rotate the source followed by  a move in the complex quaternion  plane as follows.\\
At first, the source is rotated:
\begin{eqnarray}\label{xy}
x_i^{\prime}=x_i \sin \phi _i- y_i \cos \phi _i,
\nonumber\\
y_i^{\prime}=x_i \cos \phi _i+y_i \sin \phi _i,
\end{eqnarray}
where $x_i^{\prime 2}+y_i^{\prime 2}=x_i^2+y_i^2, ~~~(i=1,2)$.\\
Then we replace ``$y^{\prime}_1\rightarrow y^{\prime}_1- \hat{j} b$" and ``$y^{\prime}_2\rightarrow y^{\prime}_2- \hat{i} a$", therefore:
\begin{eqnarray}\label{r02}
&\frac{1}{r_0^2}\rightarrow \frac{1}{r_c^2}=\frac{1}{x_1^{\prime 2}+(y_1^\prime- \hat{j} b)^2+x_2^{\prime 2}+(y_2^\prime- \hat{i} a)^2}
\nonumber\\
&~~~~~~~~~~~~~~~~~~~~~~~~~~=\frac{1}{x_1^{\prime 2}+y_1^{\prime2}-b^2-(2 y_1^\prime \hat{j} b)
+x_2^{\prime 2}+y_2^{\prime 2}-a^2-2(y_2^\prime \hat{i} a)}.
\end{eqnarray}
Using Eq. (\ref{2}) we have
\begin{eqnarray}\label{yy}
x_1^{\prime 2}+y_1^{\prime2}=(r^2+a^2)\sin ^2\theta,~~~~~y_1^\prime =r \sin \theta,
\nonumber\\
x_2^{\prime 2}+y_2^{\prime 2}=(r^2+b^2)\cos ^2\theta,~~~~~y_2^\prime =r \cos \theta,
\end{eqnarray}
Replacing Eqs. (\ref{yy}) into Eq. (\ref{r02}), leads to
\begin{eqnarray}\label{29}
&\frac{1}{r_c^2}=\frac{1}{(r^2+a^2)\sin ^2\theta-b^2- 2 \hat{j} r \sin \theta b+
(r^2+b^2)\cos ^2\theta-a^2- 2 \hat{i} r \cos \theta a
}
\nonumber\\
&=\frac{1}{r^2-a^2 \cos ^2\theta - 2 \hat{j} r \sin \theta b-
b^2\sin ^2\theta- 2 \hat{i} r \cos \theta a.
}
\end{eqnarray}
Considering the properties of  quaternions 
 ($\hat{i}  \cdot  \hat{j} =\hat{k}$ and $\hat{j} \cdot  \hat{i}=-\hat{k}$), Eq. (\ref{29}) maybe written as:
 \begin{eqnarray}
 \frac{1}{r_c^2}=\frac{1}{(r-\hat{i} a \cos \theta -\hat{j} b \sin \theta)^2}.
 \end{eqnarray}
 In this way we have explained ``$r\rightarrow r-\hat{i} a \cos \theta - \hat{j} b \sin \theta$" in five dimensions.\\
 It is important to consider the governing rules for $f(r)$ in the Ker-Schild form of metric ($g_{a b}=\eta _{ab}+f(r) \ell _a\ell _b$). In general, for finding $f(r)$, the following rules should be applied \cite{4}:
  \begin{eqnarray}
 r\rightarrow \frac{1}{2}(r+ \bar{r}) = Re r,\\
 \frac{1}{r}\rightarrow \frac{1}{2}(\frac{1}{r}+\frac{1}{\bar{r}})= \frac{Re r}{\vert r \vert ^2},\\
 r^2\rightarrow  r \bar{r}.
 \end{eqnarray}
Therefore, $\frac{m}{r_0^2}$ can be rewritten as:
 \begin{eqnarray}
 \frac{m}{r_0^2}\rightarrow \frac{m}{r^2+a^2 \cos ^2 \theta +b^2 \sin ^2 \theta}.
 \end{eqnarray}
 \\
As quaternions are not defined in higher dimensions, at this time we cannot extend our arguments to six or higher dimensions. However,  for special case of equal angular 
momenta in odd space-time dimensions, we propose a version of the NJ trick in the following section.  \\
For example in five dimensions by using Eq. (\ref{1}) and Eq. (\ref{2}), we have
\begin{equation}\label{1xy}
x _{1}^2+y _{1}^2+x _{2}^2+ y _{2} ^2=r_0 ^2,
\end{equation}
and 
\begin{equation}
 \label{2xy}
x _1^2+y _1^2+x _2^2+ y _2^2 =r^2+a^2.
\end{equation} 
Comparing Eq. (\ref{1xy}) and (\ref{2xy}) implies:
\begin{equation}
r_0^2\rightarrow r^2+a^2,
\end{equation}
It is worth nothing that this replacement works for odd higher dimensions.
\\
In the next Section, we consider the  higher-dimensional rotating black holes with equal angular momenta. 
\section{Rotating odd-dimensional black holes with equal angular momenta}\label{AppendixA}
Here, we consider the rotating black holes in higher dimensions, where all angular momenta are equal  and propose a new version of the NJ algorithm.
The focus is on the Kerr--Schild coordinate in the odd space-time dimension.
\\The  $d=2n+1$
dimensional metric $(d\geq 5)$ in the Cartesian coordinates is as follows:
\begin{eqnarray}
&ds^2=\mathsf{g}_{a b} dx^a dx^b~~~~~~~~~~~~~~~~~~~~~~~~~~~~~~~~~~~~~~~~~~~~~~~~~~~~~~~~~~~~~~~~~~~~~~~~~~~~~~~~~
\nonumber\\
&=-d\tau^2+\sum_{i=1}^{n}(dx_{i}^2+dy_{i}^2)+(1-f(r_0)) \left( \frac{\sum_{i=1}^{n}(x_{i} dx _{i}+y _{i} dy _{i})}{r_0}
+d\tau \right)^2,
\end{eqnarray}
where,
\begin{equation}
 f(r_0)=1-\frac{m}{r_0^{d-3}}.
\end{equation}
The generic form of the Kerr--Schild metric in the d dimensions is presented by \cite{Higher Dimensions ,Higher Dimensions1,Higher Dimensions2}:
\begin{eqnarray}\label{g}
\mathsf{g}_{a b}= \eta_{a b}+(1-f(r_0))(\ell_0)_a (\ell_0)_b,
\end{eqnarray}
where,
 $\eta_{a b}$ is the Minkowski metric and
 \begin{eqnarray}
&(\ell_0)_a=(1;\widehat{r_0})~~~~~~~~~~~~~~~~~~~~~~~~~~~~~~~~
\nonumber\\
&=(1;\frac{x_{1}}{r_0},\frac{y_{1}}{r_0},\cdots,\frac{x_{n}}{r_0},\frac{y_{n}}{r_0}).
\end{eqnarray}
Consider the following coordinate transformations:
\begin{eqnarray}
x_{i}=r_0 \mu _{0i} \cos \phi_{0i} ,~~~~~y_{i}=r_0 \mu _{0i} \sin \phi _{0i}, ~~~i=1...n,
\end{eqnarray}
or
\begin{eqnarray}
x_{i}+\hat{i}y_{i}=r_0 \mu _{0i} e^{\hat{i}\phi_{0i}},~~~~~~~~~~\sum _{i=1}^{n} \mu _{0i} ^2=1.
\end{eqnarray}
It can be shown that
\begin{eqnarray}\label{4.7}
\sum_{i=1}^{n}(x_{i}^2+y^2_{i})=r_0^2.
\end{eqnarray}
In this case, $(\ell_0)_a$ can be written as follows:
\begin{equation}
 (\ell_0)_a=(1;\mu_{01} \cos \phi_{01},\mu_{01} \sin \phi_{01},\cdots ,\mu_{0n} \cos \phi_{0n},\mu_{0n}  \sin \phi_{0n}).
 \end{equation}
For obtaining  the rotating black holes, we initially introduce the following coordinates
\begin{equation}
x _i+\hat{i} y _i =(r+\hat{i} a_i) \mu_i e^{\hat{i } \phi_i}, ~~~~~~\sum_{i=1}^{n}\mu_i^2=1.
\end{equation}
Therefore, we obtain the following equation:
\begin{eqnarray}\label{4.11}
\sum_{i=1}^{n}(x_{i}^2+y^2_{i})=r^2+\sum_{i=1}^{n}a_i^2 \mu_i^2.
\end{eqnarray}
It should be noted that the right hand side of Eq. (\ref{4.11}) does not depend on $\mu_i$ $(\sum_{i=1}^{n}(x_{i}^2+y^2_{i})=r^2+a^2 )$  for equal angular momenta $(a_i=a)$. However, for even dimensions ($d=2n+2$), an unpaired spatial coordinate, $z$, exists ($\sum_{i=1}^{n}\mu_i^2+\alpha^2 =1,~~ (\alpha =z/r)$). Consequently, this result is not valid for these dimensions with equal angular momenta.

 At first step, we replace $(\ell_0)_a$ as follows:
\begin{eqnarray}
(\ell_0)_a
\longrightarrow \ell _a =&&(1;\mu_1 \cos \phi_1,\mu_1 \sin \phi_1,\cdots,\mu_n \cos \phi_n,\mu_n \sin \phi_n),
\nonumber\\
=&&(1; \mathit{R}\textit{e}\Big[\frac{x_1+\hat{i} y_1}{r+\hat{i} a_1} \Big], \mathit{I}\textit{m}\Big[\frac{x _1+\hat{i} y _1}{r+\hat{i} a_1} \Big], \cdots,\mathit{R}\textit{e}\Big[\frac{x _n+\hat{i} y_n}{r+\hat{i} a_2} \Big], \mathit{I}\textit{m}\Big[\frac{x _n+\hat{i} y _n}{r+\hat{i} a_2} \Big])
\nonumber\\
=&&(1; \frac{x _1 r+y _1 a_1}{r^2+a_1^2} , \frac{y_1 r- a_1 x_1}{r^2+a_1^2},\cdots , \frac{x_n r+y _n a_n}{r^2+ a_n^2} , \frac{y _n r- a_n x _n }{r^2+ a_n^2}).
\end{eqnarray}
At the second step,  for equal angular momenta, comparing Eqs. (\ref{4.7}) and  (\ref{4.11}) implies the following replacement:
\begin{equation}
r_0^2\rightarrow r^2+a^2,
\end{equation}
 leading to the following
 transformation ($1-f(r_0)\rightarrow 1-f(r)$):
\begin{eqnarray}
\frac{m}{(r_0^2)^{\frac{d-3}{2}}}\rightarrow \frac{m}{(r^2+a^2)^{\frac{d-3}{2}}}.
\end{eqnarray}
Finally, the following  metric is obtained for the rotating black holes
\begin{eqnarray}
\mathsf{g}_{a b}= \eta_{a b}+(1-f(r))(\ell)_a (\ell)_b.
\end{eqnarray}
Hence  for equal angular momenta $a_i=a$, we have:
\begin{eqnarray}
ds^2&&=-d\tau^2+\sum_{i=1}^{n}(
dx_i^2+dy _i^2)+
\nonumber\\
&&(1-f(r)) \left(\sum_{i=1}^{n} \frac{r(x_i dx _i+y _i dy _i)-a(x _i dy_i-y _i dx _i)}{r^2+a^2}
+d\tau\right)^2,
\end{eqnarray}
where
\begin{equation}
f(r)=\frac{m}{(r^2+a^2)^{\frac{d-3}{2}}}.
\end{equation}
In this way, by using the Kerr--Schild coordinates,  the metric of rotating (2n+1)-dimensional black holes with equal angular momenta is obtained by  employing  a simple algorithm.
\section{Conclusion}\label{Conclusion}
In this paper, we proposed a simple approach for deriving rotating black holes from static ones.
The Kerr--Schild coordinates and the quaternion features are used in this approach for obtaining five-dimensional rotating black holes. This algorithm  includes two steps, i.e. replacement of the null co-vectors in the Kerr--Schild coordinates  $(\ell_0)_a \rightarrow \ell_a$ and use of quaternions for replacing $f(r_0) $ by $f(r)$.
We also explain the process of obtaining the  distance parameter replacement by using rotation and translation of resource in the potential.
At this time we cannot generalize our five dimensional  approach to higher dimensions, and it remains for future studies.
 In the second part of our paper,
 we proposed a new version of the NJ algorithm for higher odd-dimensional black holes with equal angular momenta for the first time.
 \\
 For the  future works, it is interesting  to generalize the method  proposed in this paper  to obtain metrics   of rotating black holes with unequal  angular momenta and also other tips of black holes.
%
%
%
\section{Acknowledgements}
This work has been supported financially by Research Institute for Astronomy and Astro physics of Maragha (RIAAM) under research project No.1/5750-50.
\section{Appendix A }\label{3}
 Numerical system of quaternions was used to extend complex numbers. Quaternions are generally represented as follows:
\begin{eqnarray}
A=a_1+\hat{i} a_2+\hat{j} a_3+\hat{k} a_4.
\end{eqnarray}
Furthermore, they can be defined by introducing the elements of a Hamilton basis 
 $ \hat{i}, \hat{j},\hat{ k}, $ satisfying the following rules:
 \begin{eqnarray}
&&\hat{i} ^2 = \hat{j} ^2 = \hat{k} ^2 = \hat{i}  \cdot \hat{j}  \cdot \hat{ k} = -1,\\
&& \hat{i}   \cdot \hat{j}=\hat{k}, ~~\hat{j}   \cdot \hat{i}=-\hat{k}.
 \end{eqnarray}
 The conjugate of $A$ is defined as below:
 \begin{eqnarray}
\bar{A}=a_1- \hat{i} a_2-\hat{j} a_3-\hat{k} a_4,
\end{eqnarray}
 and its norm is defined as follows:
\begin{eqnarray}
\vert A \vert ^2= A \bar{A}=a_1^2+a^2_2+a_3^2+a_4^2.
\end{eqnarray}
We use this quantity to obtain (\ref{3.10}).
\section{Appendix B}
\label{sec:2}
The Newman-Janis  method can be used to obtain rotating BTZ black holes \cite{BTZ1,BTZ2}. In \cite{BTZ},   four-dimensional transformations were used to derive rotating BTZ black hole solution from the non-rotating one, and the final three-dimensional result was presented in the equatorial plane (slice in $\theta = \frac{\pi}{2}$). We propose  a simple trick to obtain three-dimensional rotating BTZ black holes.
This simple method is based on the Kerr--Schild coordinate system.
\\
In Cartesian coordinates, the non-rotating(static) three-dimensional BTZ  metric is as follows:
\begin{eqnarray}\label{2.1}
&ds^2=\mathsf{g}_{a b} dx^a dx^b~~~~~~~~~~~~~~~~~~~~~~~~~~~~~~~~~~~~~~~~~~~~~~~~~~~~~
\nonumber\\
&~~~~~=-d\tau^2+dx_0^2+dy_0^2+(1-f(r_0))(\frac{x_0 dx_0+y_0 dy_0}{r_0}+d\tau)^2,
\end{eqnarray}
where
\begin{equation}
f(r_0)=(-M+r_0^2/\ell^2).
\end{equation}
The components of metric in (\ref{2.1})  can be written  as:
\begin{eqnarray}
\mathsf{g}_{a b}=\eta_{a b}+\big(1- f(r_0)\big) (\ell_0)_a (\ell_0)_b;
\end{eqnarray}
where
\begin{eqnarray}
&&\eta  _{a b}=diag \lbrace -1,+1,+1 \rbrace,
\\
&&(\ell_0)_a=(1;\hat{r_0})
\nonumber\\
&&~~~~~~=(1;\frac{ x_0 }{r_0},\frac{y_0}{r _0}).
\end{eqnarray}
Using the following coordinate transformation
\begin{equation}
x +\hat{i} y =r_0 e^{\hat{i} \phi_{0}},
\end{equation}
 a new form of $(\ell_0 )_a$ can be obtained as follows:
\begin{equation}
(\ell_0)_a=(1; \cos \phi_{0},\sin \phi_{0}).
\end{equation}
To derive rotating  space-time black holes from static one, the
 following coordinates are initially used
\begin{equation}
x +\hat{i}y  =(r+\hat{i}a) e^{\hat{i} \phi},
\end{equation}
where $a$ is the angular momentum parameter and transforms $(\ell_0)_a$ to $\ell_a$ as follows:
\begin{eqnarray}
(\ell_0)_a
\longrightarrow \ell _a &&=(1; \cos \phi, \sin \phi)
\nonumber\\
&&=(1; \mathit{Re}\Big[\frac{\mathrm{x}+\hat{i}y}{r+\hat{i} a} \Big], \mathit{Im}\Big[\frac{x +\hat{i}y }{r+\hat{i}a} \Big])
\nonumber\\
&&=(1; \frac{r x +a y }{r^2+a^2} , \frac{ry- a x}{r^2+a^2}).
\end{eqnarray}
Next, we replace
\begin{equation}
(-M+r_0^2/\ell^2)\longrightarrow (-M+r^2/\ell^2),
\end{equation}
the following transformation is used,
$r_0\longrightarrow r+\hat{i} a \sqrt{1-\mu ^2}$ where   $\mu$  is equal to one for BTZ black holes. Finally,  we have
\begin{eqnarray}
\mathsf{g}_{a b}=\eta_{a b}+(1-(-M+r^2/\ell^2)) (\ell)_a (\ell)_b.
\end{eqnarray}
That is, the rotating three-dimensional BTZ metric may be obtained  as follows:
\begin{eqnarray}
& ds^2  =-d \tau^2+dx^2+dy^2+
(1-f(r)) \left( \frac{r (x dx +y dy  )-a (x dy  -ydx   )}{r^2+a^2}
+d\tau \right)^2,
\nonumber\\
&f(r)=(-M+r^2/\ell^2).
\end{eqnarray}
%
%

\end{document}